\documentclass{cjpsuppl}     
\usepackage{epsfig,rotating}
\begin{document}
\title{Selected Parity Violation Experiments}
\authori{W.\, D. Ramsay}
\addressi{Department of Physics and Astronomy, University of Manitoba,
Winnipeg, MB, R3T 2N2, Canada}
\authorii{}    \addressii{}
\authoriii{}   \addressiii{}
\authoriv{}    \addressiv{}
\authorv{}     \addressv{}
\authorvi{}    \addressvi{}
\headtitle{Selected Parity Violation Experiments \ldots}
\headauthor{W.D. Ramsay}
\lastevenhead{W.D. Ramsay: Selected Parity Violation Experiments \ldots}
\pacs{11.30.Er, 24.70.+s, 25.40.Cm, 25.40.Lw, 13.60.Fz}
\keywords{parity violation, polarized beams, proton, electron, form factor}
\refnum{}
\daterec{20 October 2003;\\final version:}
\suppl{A}  \year{2003} \setcounter{page}{1}
\maketitle

\begin{abstract}

I start by reviewing existing $\vec{p}p$ measurements with particular emphasis on
the recent 221 MeV $\vec{p}p$ measurement at TRIUMF which permitted the weak
meson-nucleon coupling constants $h^{pp}_\rho$ and $h^{pp}_\omega$ to be
determined separately for the first time.  I then review $\vec{n}p$
experiments, with specific details of the $\vec{n}p \rightarrow d\gamma$
experiment now under preparation at Los Alamos National Laboratory. This
experiment will provide a clean measurement of the weak pion nucleon coupling,
$f_\pi$. Finally, I discuss $\vec{e}p$ parity violation experiments,
particularly the Gzero experiment under way at Jefferson Lab in Virginia. This
experiment will measure the weak form factors $G_E^z$ and $G_M^z$, allowing the
distribution of strange quarks in the quark sea to be determined.

\end{abstract}

\section{Introduction}

In Vancouver a popular form of Chinese luncheon is ``Dim Sum'' in which small
quantities of a large variety of foods may be tasted.  This review is a ``Dim
Sum'' of parity violation experiments.  As with a luncheon, my selection is
biased by my personal taste and experience. I start with $\vec{p}p$ parity
violation experiments, concentrating on the the TRIUMF 221 MeV $\vec{p}p$
experiment, then discuss $\vec{n}p$ parity violation experiments with details
of the Los Alamos $\vec{n}p \rightarrow d\gamma$ experiment now being
installed at LANSCE. Finally, I discuss $\vec{e}p$ parity violation
experiments, particularly the Gzero experiment at Jefferson Lab. I refer
those interested in more background to specific reviews on nucleon-nucleon
\cite{Ade85,Hae95} and $\vec{e}p$ \cite{Mckeown03} experiments.

\section{$\vec{p}p$ Experiments}

\begin{figure}
\includegraphics[height=0.8\textwidth,angle=-90]{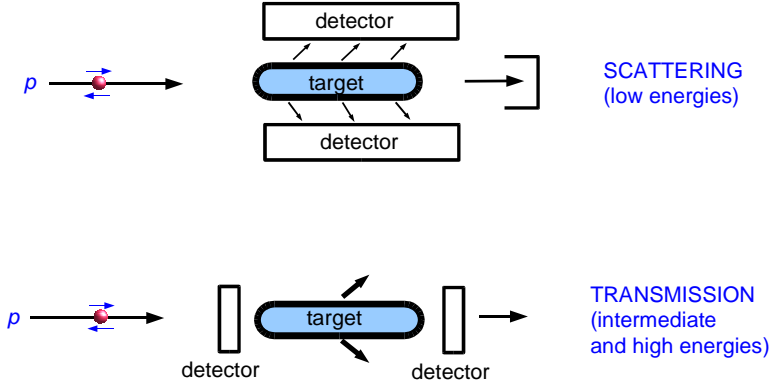}
\caption{Types of $\vec{p}p$ experiments. The low-energy experiments use
scattering geometry, while the intermediate and high-energy experiments
use transmission geometry.}
\label{pptypes}
\end{figure}

Figure \ref{pptypes} shows typical $\vec{p}p$ parity violation experiments.
They scatter a longitudinally polarized beam of protons from a hydrogen target
and measure the difference in cross section for right-handed and left-handed
proton helicities. The intermediate and high energy experiments use
transmission geometry in which the change in scattering cross section is
deduced from the change in transmission through the target. Low energy
experiments, where energy loss limits the target thickness, use scattering
geometry, in which the detectors measure the scattered protons directly.  Both
types of experiments measure the parity violating longitudinal analyzing power,
$A_z = \frac{\sigma^+ - \sigma^-}{\sigma^+ + \sigma^-}$, where $\sigma^+$ and
$\sigma^-$ are the scattering cross sections for positive and negative
helicity. 

\begin{table}[!h]
\caption{Summary of $\vec{p}p$ parity violation experiments. The long times
taken to achieve small uncertainties reflects the time taken to understand
and correct for systematic errors. In cases where authors reported both
statistical and systematic uncertainties, this table shows the quadrature
sum of the two.} \vspace{1mm}
\footnotesize
\begin{center}
\begin{tabular}{|l|l|c|r|}
\hline
  \multicolumn{1}{|c}{\bf Lab/Energy} &
  \multicolumn{1}{c}{\bf Technical Details} &
  \multicolumn{1}{c}{\bf $A_z$ ($10^{-7}$)} &
  \multicolumn{1}{c|}{\bf Where Reported}\\
\hline
Los Alamos & scattering                  & $+1 \pm 4$       & 1974 Phys. Rev. Lett. \cite{Pot74}     \\
15 MeV     & 3 atm x 38cm hydrogen gas   &                  &                                        \\
           & 4 liquid scintillators      &                  &                                        \\
\hline
           & scattering                  &  $-1.7 \pm 0.8$  & 1978 Argonne Conference \cite{Nag79}   \\
           & 6.9 atm hydrogen gas        &                  &                                        \\
           & 4 plastic scintillators     &                  &                                        \\
\hline
Texas A\&M & scattering                  & $-4.6 \pm 2.6$   & 1983 Florence Conference \cite{Tan83}  \\
47 MeV     & 39 atm x 42cm hydrogen gas  &                  &                                        \\
           & 4 plastic scintillators     &                  &                                        \\
\hline
Berkeley   & scattering                  & $-1.3 \pm 1.1$   & 1980 Santa Fe Conference \cite{vonR80} \\
46 MeV     & 80 atm hydrogen gas target  &                  &                                        \\
           & He ion chamber around target& $-1.63 \pm 1.03$ & 1985 Osaka Conference \cite{vonR85}    \\
\hline
SIN (PSI)  & scattering                  & $-3.2 \pm 1.1$   & 1980 Phys. Rev. Lett. \cite{Bal80}     \\
45 MeV     & 100 atm hydrogen gas        &                  &                                        \\
           & annular ion chamber         & $-2.32 \pm 0.89$ & 1984 Phys. Rev. D. \cite{Bal84}        \\
           &                             &                  &                                        \\
           &                             & $-1.50 \pm 0.22$ & 1987 Phys. Rev. Lett. \cite{Kist87}    \\
\hline
Los Alamos & transmission                & $+2.4 \pm 1.1$   & 1986 Phys. Rev. Lett. \cite{Yua86}     \\
800 MeV    & 1 m liquid hydrogen gas     &                  &                                        \\
           & ion chambers                &                  &                                        \\
\hline
Bonn       & scattering                  & $-1.5 \pm 1.1$   & 1991 Phys. Lett. B \cite{Evers91}      \\
13.6 MeV   & 15 atm hydrogen gas         &                  &                                        \\
           & hydrogen ion chambers       & $-0.93 \pm 0.21$ & 1994 private communication \cite{Evers94}\\
\hline
TRIUMF     & transmission                & $+0.84 \pm 0.34$ & 2001 Phys. Rev. Lett. \cite{Berd01b}   \\
221 MeV    & 40 cm liquid hydrogen       &                  &                                        \\
           & hydrogen ion chambers       &                  &                                        \\
\hline
Argonne ZGS& transmission                & $+26.5 \pm 7.0$ & 1986 Phys. Rev. Lett. \cite{Loc84}      \\
5130 MeV   & 81 cm water target          &                  &                                        \\
           & ion chambers                &                  &                                        \\
           & and scintillators           &                  &                                        \\
\hline

\end{tabular}
\vspace{-1mm}
\end{center}
\label{ppexp}
\end{table}

A roughly historical summary of $\vec{p}p$ parity violation experiments is
given in Table 1. The long time taken to acquire measurements at a
reasonable selection of energies and with small experimental uncertainties
reflects the technical difficulty of these measurements.  Running time is
dominated by the time required to understand, and correct for, the various
sources of systematic error. The time required to get the desired
statistical precision is normally small by comparison.

\begin{table}[h]
\caption{Overall corrections for systematic errors in the TRIUMF parity
violation experiment. The table shows the average value of each coherent
modulation, the net correction made for this modulation, and the uncertainty
resulting from applying the correction.}  \vspace{1mm}
\small
\begin{center}
\begin{tabular}{|l|c|r|}
\hline
 Property & Average Value & $10^7 \Delta A_z$  \\
\hline
 $A_z^{uncorrected} (10^{-7})$ & ~~~~~~~$1.68 \pm 0.29(stat.)$ & \\
 $y*P_x (\mu m)$ & $-0.1 \pm 0.0$ & $-0.01 \pm 0.01$ \\
 $x*P_y (\mu m)$ & $-0.1 \pm 0.0$ & $0.01 \pm 0.03$ \\
 $\langle yP_x \rangle (\mu m)$ & $1.1 \pm 0.4$ & $0.11 \pm 0.01$ \\
 $\langle xP_y \rangle (\mu m)$ & $-2.1 \pm 0.4$ & $0.54 \pm 0.06$ \\
 $\Delta I/I (ppm)$ & $15 \pm 1$ & $0.19 \pm 0.02$ \\
 $position + size$ &           & $     0  \pm 0.10$ \\
 $\Delta E(meV at\, OPPIS)$&   7--15      & $  0.0  \pm 0.12$ \\
 electronic crosstalk &     & $ 0.0 \pm 0.04$ \\
 Total & & $0.84 \pm 0.17 (syst.)$ \\
\hline\hline
 $A_z^{corr} (10^{-7})$ &
\multicolumn{2}{c|}{$0.84 \pm 0.29(stat.) \pm 0.17(syst.) $} \\
\hline

\end{tabular}
\vspace{-1mm}
\end{center}
\label{corr}
\end{table}

\begin{figure}[t]
\begin{center}
\includegraphics[height=0.80\textwidth,angle=90]{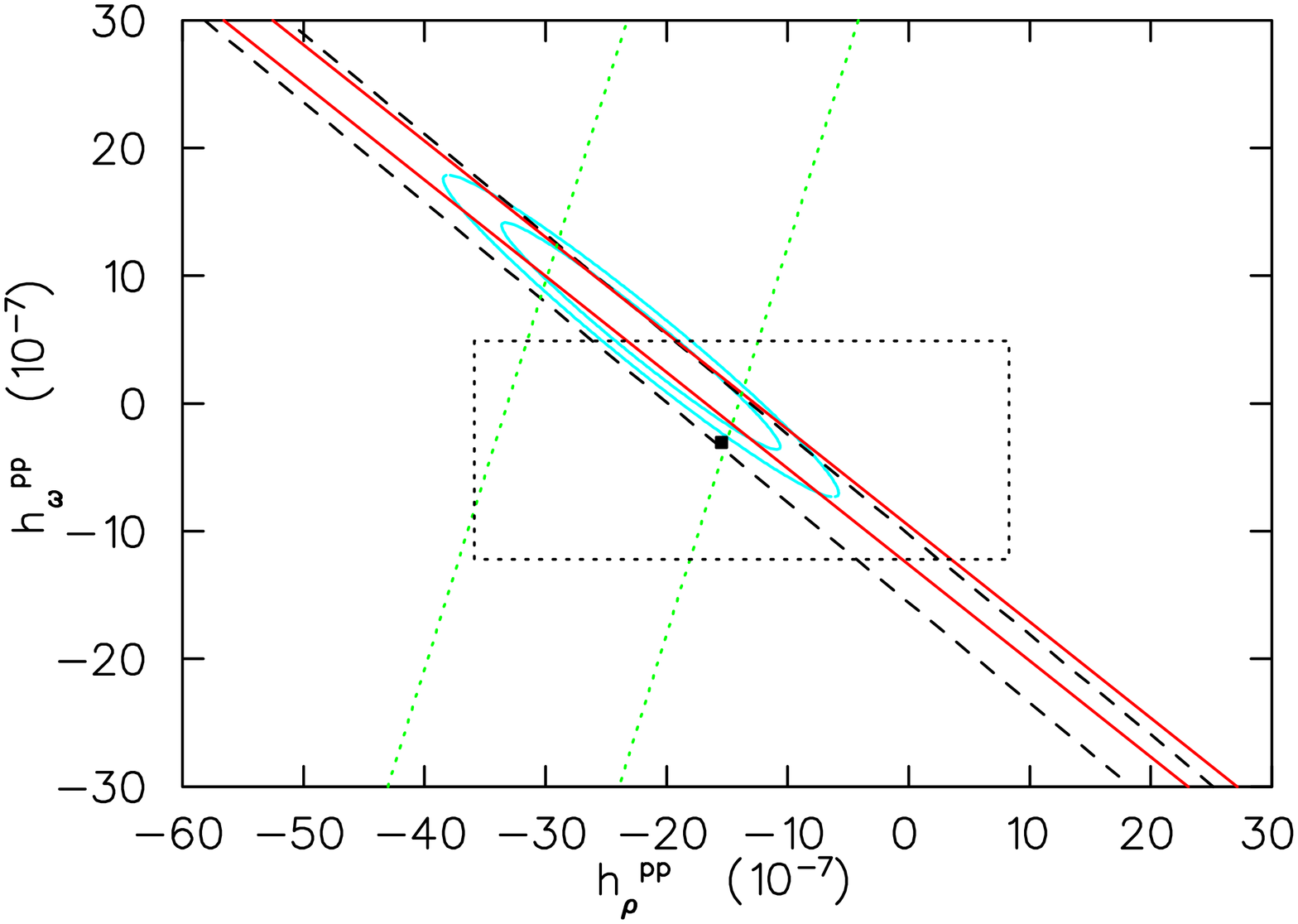}
\end{center}
\caption{Constraints on the weak meson-nucleon couplings imposed by experiments
in the energy range where the meson exchange model is normally used. The bands
are based on calculations by Carlson {\em et al.} \cite{Carl01} using the AV18
potential \cite{Wirin95} and CD-Bonn strong couplings \cite{Mach01}. The
contours are 68\% and 90\% C.L. (Figure modified from \cite{Berd01b})}.
\label{limits}
\end{figure}

The TRIUMF $pp$ experiment \cite{Berd01b} is a transmission experiment as shown
in the lower panel of figure \ref{pptypes}. A 221 MeV longitudinally polarized
proton beam was passed through a 400 mm long liquid hydrogen target, which
scattered about 4\% of the beam. Hydrogen filled ion chambers located upstream
and downstream of the target measured the change in transmission when the spin
of the incident protons was flipped from right-handed to left-handed. Although
a very good optically pumped polarized ion source \cite{Zelen96,Zelen97,Lev95}
was used that minimized the changes in beam properties other than helicity,
other beam properties still changed very slightly. These helicity-correlated
beam property changes caused a systematic shift in the $A_z$ distribution, and
corrections must be made. To do this, the TRIUMF group continuously measured
the helicity correlated changes in beam properties and made corrections based
on the sensitivities determined in separate control measurements. All the
corrections are summarized in Table 2. The importance of accurate corrections
is apparent when one notes that the measured raw $A_z$ actually came half from
true parity violation and half from false effects.

Because the range of the W and Z bosons carrying the weak force is so small
($\sim 0.002 \:fm$), the low and intermediate energy $\vec{p}p$ results are
normally interpreted using meson exchange models, and parameterized in terms of
a set of $\pi$, $\rho$, and $\omega$ weak meson-nucleon coupling constants. 
$\vec{p}p$ experiments are sensitive to the combinations $h_{\rho}^{pp} =
h_{\rho}^{(0)}+h_{\rho}^{(1)}+ \frac{1}{\sqrt{6}} h_{\rho}^{(2)}$ and
$h_{\omega}^{pp}=h_{\omega}^{(0)}+h_{\omega}^{(1)}$. Where the subscript
denotes the exchanged meson and the superscript the isospin change.

The value of the couplings can be extracted from the experiments by assuming a
realistic model for the strong interaction and adjusting the weak couplings to
fit the data. Using the AV18 strong potential \cite{Wirin95} and CD-Bonn
values for the strong couplings \cite{Mach01}, Carlson {\em et al.}
\cite{Carl01} calculate that
\begin{eqnarray*}
A_z(13.6 MeV) & = &  0.059h_{\rho}^{pp}+0.075h_{\omega}^{pp} \\
A_z(45 MeV)   & = &  0.10h_{\rho}^{pp}+0.14h_{\omega}^{pp} \\
A_z(225 MeV)  & = & -0.038h_{\rho}^{pp}+0.010h_{\omega}^{pp}
\end{eqnarray*}
where the energies correspond to the most accurate measurements over the low
and intermediate energy range \cite{Kist87,Evers94,Berd01b}. These constraints
are shown graphically in Fig. \ref{limits}. Note that the low energy results
scale as $\sqrt{E}$ and hence constrain essentially the same linear combination
of $h_{\rho}^{pp}$ and $h_{\omega}^{pp}$. It was only when the TRIUMF result
became available that $h_{\rho}^{pp}$ and $h_{\omega}^{pp}$ could be separately
determined. By adjusting the couplings for the best fit to the data, Carlson
{\em et al.} \cite{Carl01} estimate $h_{\rho}^{pp} = -22.3 \times 10^{-7}$ and
$h_{\omega}^{pp} = 5.17 \times 10^{-7}$ compared to the DDH \cite{DDH80}
theoretical ``best guess'' values of $h_{\rho}^{pp} = -15.5 \times 10^{-7}$ and
$h_{\omega}^{pp} = 3.0 \times 10^{-7}$

\section{$\vec{n}p\rightarrow d\gamma$ Experiments}

\begin{figure}
\includegraphics[width=\textwidth]{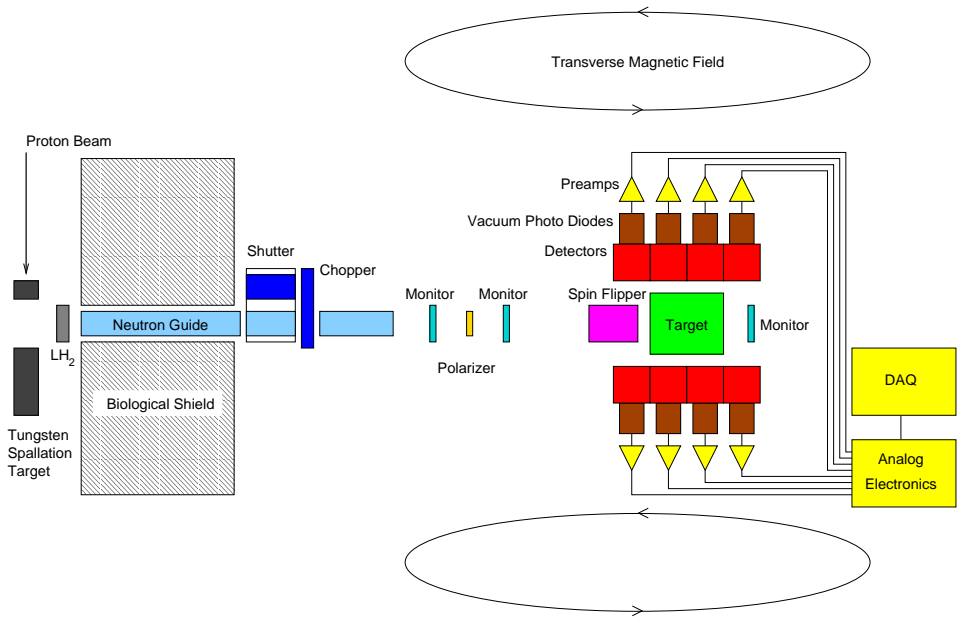}
\caption{Layout of apparatus for the $\vec{n}p \rightarrow d\gamma$
experiment at LANSCE.}
\label{npdgexp}
\end{figure}

Unlike the $\vec{p}p$ experiments just discussed, which are sensitive to $\rho$
and $\omega$ exchange,  $\vec{n}p \rightarrow d\gamma$ experiments are
sensitive almost exclusively to pion exchange, and measure the weak
pion-nucleon coupling, $f_{\pi}$.  In an $\vec{n}p \rightarrow d\gamma$
experiment, the incident cold neutrons are polarized vertically and the gamma
rays produced by neutron capture in the hydrogen target are expected to be
emitted slightly more in the direction opposite to the neutron spin. The
up-down asymmetry $A_{\gamma} \approx -0.11 f_{\pi}$ provides a clean measure
of $f_{\pi}$ \footnote{Some authors quote
$H_{\pi}=f_{\pi}\frac{g_{\pi}}{\sqrt{32}}$, where $g_{\pi}$ is the strong
pion-nucleon coupling.} free of nuclear structure uncertainties \cite{Snow00}.
Previous measurements at ILL Grenoble gave $A_{\gamma}=(6\pm 21) \times
10^{-8}$ \cite{Cav77} and $A_{\gamma}=(-1.5\pm 4.8) \times 10^{-8}$
\cite{Alb88}, but neither result was accurate enough to impose a significant
constraint.

An experiment is now being prepared at Los Alamos to measure the gamma ray
asymmetry in  $\vec{n}p \rightarrow d\gamma$ with an uncertainty of $\pm 0.5
\times 10^{-8}$ \cite{Snow00}. The expected asymmetry is $A_{\gamma} \approx -5
\times 10^{-8}$. The apparatus is shown schematically in Fig. \ref{npdgexp}. 
Neutrons are produced by an 800 MeV proton beam incident on a tungsten
spallation target. The neutrons are cooled in a liquid hydrogen moderator and
transported to the experiment through a super mirror neutron guide. The
neutrons are polarized vertically by a polarized $^3He$ spin filter then
captured in a liquid para-hydrogen target. The gamma asymmetry is measured by
an array of 48 15 $\times$ 15 cm$^2$ CsI(Tl) detectors surrounding the
target.  

The neutron beam is pulsed at 20 Hz, so the energy of the neutrons arriving at
the experiment after the 22 m flight path can be determined by time of flight.
An RF spin flipper provides a method of rapid spin reversal to control
systematic errors.  Systematic errors can be further understood and controlled
by reversing the $^3He$ cell direction or the direction of the overall vertical
10 gauss guide field.  In addition, different systematic errors have different
dependences on time of flight.

The beamline, FP12, is now complete and the experimental cave is scheduled for
completion in fall, 2003.  Commissioning runs will follow, with the first
production data taking anticipated in late 2004 and 2005.

\begin{figure}[!h]
\includegraphics[width=50mm, height=125mm, angle=90]{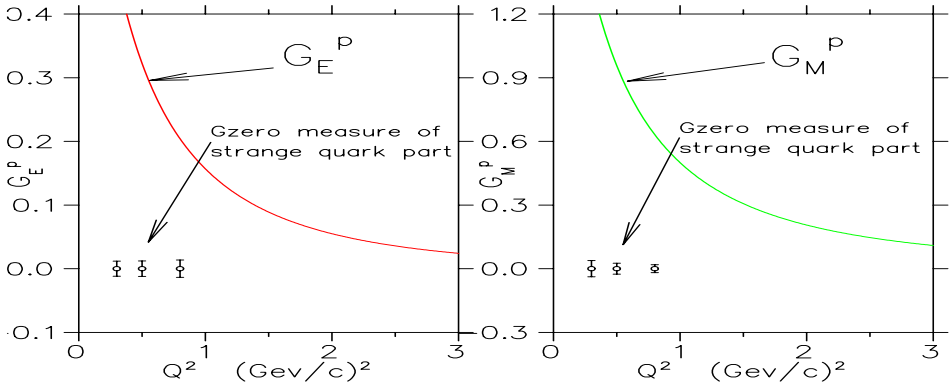}
\caption{The proton form factors $G^p_E$ and $G^p_M$ \cite{Brash02} are the sum
of contributions from up down and strange quarks. The points with error bars
show the anticipated uncertainty in the Gzero measurement of the strange
quark part.}
\label{gzero-error}
\end{figure}

\section{$\vec{e}p$ Experiments -- The Gzero Experiment}

The Gzero experiment \cite{Beck00} at Jefferson Lab scatters a longitudinally
polarized electron beam from a 200 mm liquid hydrogen target, and measures the
parity-violating longitudinal analyzing power $A_z = \left[ \frac{1}{P}
\right]\frac{\sigma^R - \sigma^L}{\sigma^R + \sigma^L}$ where $\sigma^R$ and
$\sigma^L$ are the cross sections for right-handed and left-handed electrons,
and $P$ is the beam polarization. $A_z$ values ranging from -3 to -35 ppm are
predicted. By measuring this quantity at a range of angles and momentum
transfers, the experiment will determine the {\em weak} charge and magnetic
form factors $G^Z_{E,M}(Q^2)$ (essentially the Fourier transforms of the
spatial distributions). Because the weak charges of the quarks are different
than the electromagnetic charges (Table \ref{quarks}), one can combine these
weak form factors with the previously measured {\em electromagnetic} form
factors $G^{p,\gamma}_{E,M}$ of the proton and $G^{n,\gamma}_{E,M}$ of the
neutron and extract the strange quark form factors
\begin{eqnarray*}
G^{s,p}_{E,M} &=& (1-4\sin^2{\theta_W})G^{p,\gamma}_{E,M}
-G^{n,\gamma}_{E,M} -G^{p,Z}_{E,M} ,
\end{eqnarray*}
where $\theta_W$ is the weak mixing angle and $G^{p,Z}_{E,M}$ are the proton
electroweak form factors to be measured by Gzero.
Figure \ref{gzero-error} shows the proton form factors $G^p_E$ and
$G^p_{M}$ taken from a fit to the existing parity conserving data
\cite{Brash02}. At low $Q^2$, where the effective wavelength of the virtual
photon probe is very long, the form factors are simply the proton charge and
magnetic moment, 1.00 and 2.79. These form factors are the sum of contributions
from the different quarks; the points on the graph show the expected error in
the Gzero measurement of the strange quark part. 

\begin{table}[!h]
\caption{Electroweak couplings of up, down, and strange quarks} \vspace{1mm}
\begin{center}
\begin{tabular}{|c|c|c|}
\hline\hline
   quark    &    electric charge    &    weak charge        \\
\hline
      u     &   $+\frac{2}{3}$   &  $+1-\frac{8}{3}\sin^2{\theta_W}$   \\
      d     &   $-\frac{1}{3}$   &  $-1+\frac{4}{3}\sin^2{\theta_W}$  \\
      s     &   $-\frac{1}{3}$   &  $-1+\frac{4}{3}\sin^2{\theta_W}$  \\
\hline\hline

\end{tabular}
\vspace{-1mm}
\end{center}
\label{quarks}
\end{table}

The charge and magnetic form factors can be separated by measuring at forward
and backward angles. In each configuration, several values of momentum
transfer, $Q^2$, in the range $0.1<Q^2<1.0$ $(GeV/c)^2$ will be measured. Do do
this, the scattered particles pass through an 8-sector superconducting magnetic
spectrometer and are detected by an array of plastic scintillators arranged in
contours of constant $Q^2$. In the forward  configuration protons are detected
at $\theta_p = 70^\circ \pm 10^\circ$ (or $\theta_e = 11^\circ \pm 4^\circ$).
Over this angular range there is a strong dependence of $Q^2$ on scattering
angle, and only one beam energy of 3 GeV is required for all $Q^2$. In  the
backward configuration, electrons are detected at $\theta_e = 110^\circ \pm
10^\circ$. In this case, $Q^2$ is only weakly dependent on scattering angle,
and the beam energy must be changed for each $Q^2$.  Beam energies of 424, 585,
and 799 MeV, corresponding to $Q^2$ of 0.3, 0.5, and 0.8  $(GeV/c)^2$ are
presently planned for the backward angles.

To extract $G^Z_{E,M}$ a correction must be made for the small contribution of
the axial form factor, $G^e_A$, to the measured $A_z$.  Gzero will determine
this experimentally by measuring quasi-elastic back angle scattering from
deuterium.  By measuring at at both forward and backward angles and with both
hydrogen and deuterium targets, Gzero will be able to determine $G^s_E$,
$G^s_M$, and $G^s_A$ separately. As shown in Table \ref{epexp}, other
$\vec{e}p$ experiments have measured, or are planning to measure, various
linear combinations of the form factors. 

\begin{table}[!h]
\caption{Comparison of $\vec{e}p$ parity violation experiments. Depending on
the scattering angle and momentum transfer, the experiments are sensitive to
different linear combinations of form factors}
\vspace{1mm}
\footnotesize
\begin{center}
\begin{tabular}{|l|c|c|c|c|c|r|}
\hline
  \multicolumn{1}{|c}{\bf Experiment} &
  \multicolumn{1}{c}{\bf $E_{beam}$} &
  \multicolumn{1}{c}{\bf $I_{beam}$} &
  \multicolumn{1}{c}{\bf $\theta_e$ (deg)}& 
  \multicolumn{1}{c}{\bf $Q^2$}&
  \multicolumn{1}{c}{\bf Target}&
  \multicolumn{1}{c|}{\bf Observable}\\
\hline
SAMPLE \cite{Hasty00,Spayde00}
             & 200 MeV  & pulsed    & 130--160 &    0.1      & LH$_2$ & $G_M^s+0.44G_A^s$ \\ 
(MIT-Bates)  &          & (2.7 mA   &          &             &        &                   \\
             &          &   peak)   &          &             & LD$_2$ & $G_M^s+2.37G_A^s$ \\
\hline
HAPPEX \cite{Aniol99,Aniol01}
             & 3.3 GeV  & 35 $\mu$A &   12.3   &   0.477     & LH$_2$ & $G_E^s+0.39G_M^s$ \\ 
(Jlab Hall-A)&          &           &          &             &        &                   \\
HAPPEX-2 \cite{Kumar99}    
             &          &           &    6     &   0.1       & LD$_2$ & $G_E^s+0.08G_M^s$ \\
He\cite{Armstrong00}
             &          &           &    6     &   0.1       & He     & $G_E^s$           \\
\hline 
PVA4 \cite{Harrach93,Mass99,Maas03}
             & 854 MeV  & 20 $\mu$A &   35     &   0.225     & LH$_2$ & $G_E^s+0.22G_M^s$ \\ 
(MAMI-Mainz) &          &           &          &   0.11      &        & $G_E^s+0.10G_M^s$ \\
\hline
G-zero \cite{Beck00,Batigne03}
             &  3 GeV   & 35 $\mu$A &  6-20    &   0.1-1.0   & LH$_2$ &              \\      
(Jlab Hall-C)& 424 MeV  &           & 100--120 &    0.3      & LH$_2$ &              \\
             & 576 MeV  &           &          &    0.5      &        & measurements   \\
	     & 799 MeV  &           &          &    0.8      &        & together give  \\
             & 424 MeV  &           & 100--120 &    0.3      & LD$_2$ &  $G_E^s, G_M^S and G_A^s$ \\
             & 576 MeV  &           &          &    0.5      &        &                   \\
	     & 799 MeV  &           &          &    0.8      &        &                    \\
	     
\hline
\end{tabular}
\vspace{-1mm}
\end{center}
\label{epexp}
\end{table}

The Gzero experiment completed a successful commissioning run of the forward
angle configuration in fall 2002 and January 2003 and all major systems are now
fully operational. Running will continue with an engineering run October to
December, 2003, and production running is scheduled to start in 2004.

\section{summary}

Parity violation experiments provide a way to study effects of the weak
interaction in the presence of the much stronger electromagnetic and strong
nuclear interactions. The polarized beam experiments I have described use
similar experimental techniques and face similar problems controlling
systematic errors. The physics addressed by these experiments can, however be
quite diverse. $\vec{n}p$ experiments constrain the weak pion-nucleon coupling
constant, $f_\pi$. $\vec{p}p$ parity violation experiments are sensitive to the
shorter range part of the nucleon-nucleon force and constrain the combinations
$h_{\rho}^{pp} = h_{\rho}^{(0)}+h_{\rho}^{(1)}+ \frac{1}{\sqrt{6}}
h_{\rho}^{(2)}$ and $h_{\omega}^{pp}=h_{\omega}^{(0)}+h_{\omega}^{(1)}$.
Finally, $\vec{e}p$ parity violation experiments, such as the Jlab Gzero
experiment, offer the opportunity to measure the contribution of strange
quark-antiquark pairs to the proton charge and magnetism.


\begin{thebibliography}{}

\bibitem{Ade85} E.G. Adelberger and W.C. Haxton, Ann. Rev. Nucl. Part.
Sci. {\bf 35}, 501 (1985).

\bibitem{Hae95} W. Haeberli and Barry R. Holstein, in {\em Symmetries
and Fundamental Interactions in Nuclei}, edited by W.C. Haxton and
E.M. Henley, (World Scientific, 1995) p. 17.

\bibitem{Mckeown03} R.D. McKeown and M.J. Ramsey-Musolf, Mod. Phys. Lett. A
{\bf 18}, 75 (2003); hep-ph/0203011 .

\bibitem{Pot74} J.M. Potter {\em et al.}, Phys. Rev. Lett. {\bf 33}, 1307
(1974).

\bibitem{Nag79} D.E. Nagle {\em et al.}, in {\em Proceedings of the 3rd
International Conference on High Energy Beams and Polarized Targets}
(Argonne, 1978), edited by L.H. Thomas, AIP Conference Proceedings 51,
New York 1979, p. 224.

\bibitem{Tan83} D.M. Tanner {\em et al.}, in {\em Proceedings of the
International Conference on Nuclear Physics} (Florence, 1983),
(Typographia, Bologna, 1983), p. 697.

\bibitem{vonR80} P. von Rossen {\em et al.}, in {\em Proceedings of the
5th International Symposium on Polarization Phenomena in Nuclear Physics}
(Santa Fe, 1980), edited by G.G. Ohlsen {\em et al.}, AIP Conference
Proceedings 69, New York, 1981, p. 1442.

\bibitem{vonR85} P. von Rossen {\em et al.}, in {\em Proceedings of the
6th International Symposium on Polarization Phenomena in Nuclear Physics}
(Osaka, 1985), J. Phys. Soc. Japan {\bf 55}, Suppl. p. 1016 (1986).

\bibitem{Bal80} R. Balzer {\em et al.}, Phys. Rev. Lett. {\bf 44}, 699
(1980).

\bibitem{Bal84} R. Balzer {\em et al.}, Phys. Rev. C {\bf 30}, 1409
(1984).

\bibitem{Kist87} S. Kistryn {\em et al.}, Phys. Rev. Lett. {\bf 58}, 1616
(1987).

\bibitem{Yua86} V. Yuan {\em et al.}, Phys. Rev. Lett. {\bf 57}, 1680
(1986).

\bibitem{Evers91} P.D. Eversheim {\em et al.}, Phys. Lett. B {\bf 256}, 11
(1991)

\bibitem{Evers94} P.D. Eversheim, private communication (1994).

\bibitem{Berd01b} A.R. Berdoz {\em et al.}, Phys. Rev. Lett. {\bf 87},
272301 (2001).

\bibitem{Loc84} N. Lockyer {\em et al.}, Phys. Rev. D {\bf 30}, 860
(1984).

\bibitem{Zelen96} A.N. Zelenski {\em et al.}, in {\em Proceedings of the
12th International Symposium on High Energy Spin Physics (SPIN96)}, edited
by C.W. de Jager {\em et al.}, (World Scientific, Amsterdam, 1997), p. 637.

\bibitem{Zelen97} A.N. Zelenski {\em et al.}, in {\em Proceedings of the
6th Conference on  Intersections Between Particle and Nuclear
Physics}, edited by T.W. Donnelly, AIP Conference Proceedings 412,
New York, 1997, p.328.

\bibitem{Lev95} C.D.P. Levy {\em et al.}, in {\em Proceedings of the
International Workshop on Polarized Beams and Polarized Gas Targets}
(Cologne, 1995), edited by H.P. gen. Schieck and L. Sydow (World
Scientific, Singapore, 1996), p. 120; A.N. Zelenski, {\em ibid.}, p. 111.

\bibitem{Wirin95} R.B. Wiringa, {\em et al.}, Phys. Rev. C {\bf 51}, 38
(1995).

\bibitem{Mach01} R. Machleidt, Phys. Rev. C {\bf 63}, 24001 (2001).

\bibitem{Carl01} J.A. Carlson, R. Schiavilla, V.R. Brown, and B.F.
Gibson, Phys. Rev. C {\bf 65}, 035502, (2002); R. Schiavilla, private
communication (2001).

\bibitem{DDH80} B. Desplanques, J.F. Donoghue, and B.R. Holstein,
Ann. Phys.(N.Y.) {\bf 124}, 449 (1980).

\bibitem{Snow00} W.M. Snow {\em et al.}, Nucl. Inst. Meth. A {\bf 440},
729 (2000).

\bibitem{Cav77} J.F. Caviagnac {\em et al.}, Phys. Lett. B {\bf 67},
148 (1977).

\bibitem{Alb88} J. Alberi {\em et al.}, Can. J. Phys. {\bf 66}, 542
(1988).

\bibitem{Beck00} D. Beck, spokesperson, Jefferson Lab proposal E00-006
(2000).

\bibitem{Brash02} E.J. Brash {\em et al.}, Phys. Rev. C {\bf 65}, 05100 (2002).

\bibitem{Hasty00} R. Hasty {\em et al.}, Science {\bf 290}, 2117 (2000).

\bibitem{Spayde00} D. Spayde {\em et al.}, Phys Rev. Lett. {\bf 84}, 1106
(2000).

\bibitem{Aniol99} K.A. Aniol {\em et al.}, Phys. Rev. Lett. {\bf 82}, 1096
(1999).

\bibitem{Aniol01} K.A. Aniol {\em et al.}, Phys. Lett. B{\bf 509}, 211 (2001)
(1999).

\bibitem{Kumar99} K. Kumar, Jefferson Lab proposal E99-115 (1999)

\bibitem{Armstrong00} D. Armstrong, Jefferson Lab proposal E00-114 (2000).

\bibitem{Harrach93} D. von Harrach, spokesperson, F.E. Maas, contact, MAMI
 experiment A4-01-93 (1993).

\bibitem{Mass99} F.E. Mass {\em et al.}, in {\em Proceedings of Parity
Violations in Atoms and Polarized Electron Scattering} (Paris, 1997), edited
by B. Frois and M.A. Bouchiat (World Scientific, Singapore, 1999).

\bibitem{Maas03} F.E. Maas, Eur. Phys. J. A {\bf 17}, 339 (2003).

\bibitem{Batigne03} G. Batigne in {\em Proceedings of the 4th International
Conference on Perspectives in Hadronic Physics} (Trieste, 2003), G0 report
G0-03-075, (this report and others are available from http://www.npl.uiuc.edu/exp/G0/docs/ ).

\end{thebibliography}
\end{document}